\documentclass[11pt]{article}

\usepackage[preprint]{acl}

\usepackage{times}
\usepackage{latexsym}
\usepackage{amsmath}
\usepackage{amssymb}
\usepackage{booktabs}
\usepackage{tabularx}
\usepackage{adjustbox}

\usepackage[T1]{fontenc}

\usepackage[utf8]{inputenc}

\usepackage{microtype}

\usepackage{inconsolata}

\usepackage{graphicx}
\usepackage[table]{xcolor}
\usepackage{tabularx,array,booktabs,xcolor}
\newcolumntype{L}{>{\raggedright\arraybackslash}X}
\newcolumntype{Y}{>{\centering\arraybackslash}X}
\newcolumntype{W}{>{\raggedright\arraybackslash}X}
\newcommand{\librispeech}{\textsc{LibriSpeech}}
\newcommand{\voxceleb}{\textsc{Voxceleb}}
\newcommand{\expresso}{\textsc{Expresso}}

\newcommand{\method}{TASLA}

\newcommand{\myfig}[1]{Figure \ref{#1}}
\newcommand{\mytable}[1]{Table \ref{#1}}

\newcommand{\mysecref}[1]{Section \ref{#1}}

\title{TASLA: Text-Aligned Speech Tokens with Multiple Layer-Aggregation}

\author{
  Ming-Hao Hsu\textsuperscript{1} \quad
  Liang-Hsuan Tseng\textsuperscript{2} \quad
  Hung-yi Lee\textsuperscript{2} \quad
  Zhizheng Wu\textsuperscript{1} \\
  \textsuperscript{1}The Chinese University of Hong Kong, Shenzhen \\
  \textsuperscript{2}National Taiwan University \\
  \texttt{hsuminghao1006@gamil.com}, \texttt{wuzhizheng@cuhk.edu.cn} \\
  \texttt{f11921067@ntu.edu.tw}, \texttt{hungyilee@ntu.edu.tw}
}

\begin{document}
\maketitle
\begin{abstract}
We propose \textbf{T}ext-\textbf{A}ligned \textbf{S}peech Tokens with Multiple \textbf{L}ayer-\textbf{A}ggregation (TASLA), which is a text‑aligned speech tokenization framework that aims to address the problem that under a low-frame-rate and text-aligned regime, single-source speech tokens may lose acoustic details during reconstruction. 
On the other hand, this paper further explains how different encoder layers collaborate to capture comprehensive acoustic features for tokenization.
Previous work, TASTE, proposed the text-aligned speech tokenization framework, which is a LM-friendly architecture, but struggles to capture acoustic details.
We address this trade‑off with two components: Multi‑Layer Dynamic Attention (MLDA), which lets each text position adaptively mix shallow/deep features from a frozen speech encoder, and Finite Scalar Quantization (FSQ), a simple per‑dimension discretization with smooth optimization.
At about 2.62 Hz (tokens/s), TASLA consistently improves prosody and achieves competitive quality over TASTE on in-domain (\librispeech{}) and OOD (\expresso{}, \voxceleb{}) sets.
We further demonstrate that dynamic layer mixing is correlated with spectral flux and explains why MLDA preserves prosody under a low frame rate with extreme feature compression.
\end{abstract}
\section{Introduction}
Large language models (LLMs) have recently transformed text understanding and generation.
To pursue a more natural interaction mode with LMs, researchers have started to research Spoken Language Models (SLM)~\cite{wang2023neuralcodeclanguagemodels, taste, kimiteam2025kimiaudiotechnicalreport, hassid2024textuallypretrainedspeechlanguage, nguyen2024spiritlminterleavedspoken}, where models process and generate speech as a first-class modality.
If we want to use voice mode to interact with the LMs, one option is a cascade pipeline that first generates text with an LM and then converts text to speech using a TTS model~\cite{DBLP:journals/corr/abs-2507-16835, DBLP:conf/acl/ShikharKMLKA0C25}.
In contrast, SLM offers a better approach, as it captures acoustic details from context and generates speech that more naturally aligns with the context~\cite{DBLP:journals/corr/abs-2504-08528, DBLP:journals/corr/abs-2502-06490}.
The key to enabling SLMs to have such an ability is speech tokenization.
Speech tokenization involves mapping continuous waveforms into discrete tokens that are compact, learnable, and acoustically informative, enabling a vocoder to reconstruct speech.

However, there are two practical problems with joint modeling of text tokens and speech tokens for an SLM.
The first problem is the trade-off between acoustic richness and LM-friendliness.
Neural speech tokenizers that generate more tokens per second could provide fine-grained acoustic details since they use a shorter time frame for speech reconstruction, and these speech tokenizers often capture only acoustic details without text information~\cite{DBLP:journals/jstsp/LiuXYWWP24}.
Therefore, they could generate very high-quality speech, but their speech token sequence is often very long~\cite{audiolm, valle}, which is not suitable for transformer-based architectures if we want to generate longer speech.

The second problem is the speech token sequence length mismatch with the text sequence length.
Standard codecs operate at fixed frame rates that yield token streams far longer than text (e.g., $\sim$12.5--50 speech Hz vs.\ $\sim$2--3 text Hz), making joint speech-text modeling harder~\cite{encodec, speechtokenizer}.
Making the LMs need to learn two modalities separately, which also increases the difficulty of the training process.
However, aggressive sequence compression helps alignment but risks losing prosodic detail.
Prior work addresses the mismatch by aligning speech tokens to text via an attention mechanism during tokenization, enabling straightforward joint modeling at very low bitrates~\cite{taste}. 

However, under strong compression of the token length, it still shows limitations in preserving fine-grained acoustics and in out-of-domain generalization. 
To overcome these problems and find a balance between bitrate and token sequence length.
We propose \textbf{TASLA}, a text-aligned speech tokenization framework that preserves acoustic detail at text length.
TASLA introduces Multi-Layer Dynamic Attention (MLDA), which enables text tokens to query a frozen speech encoder and adaptively combine shallow and deep representations, allowing each text position to gather the most predictive acoustic evidence for content, prosody, and speaker cues.
To discretize reliably, we replace Residual Vector Quantization (RVQ) with Finite Scalar Quantization (FSQ)~\cite{fsq}, which provides smooth optimization and resilience to codebook pathologies.

This work shows that \emph{dynamic, per-token fusion across encoder depth} is the missing piece for text-aligned speech tokens: MLDA lets each word position select the most predictive mixture of shallow and deep features, thereby retaining prosody and speaker cues at a \emph{text-length} rate.
Under rate parity with prior text-aligned tokenizers, MLDA is consistently comparable in quality and outperforms on prosody metrics in both in-domain and OOD settings, and yields better spoken continuation than strong text-only and speech-token baselines, demonstrating that adaptive layer mixing effectively narrows the gap between alignment convenience and acoustic expressivity.
\section{Related Work}

\paragraph{Neural Audio Codecs.}
CNN-based neural codecs typically use RVQ and multi-scale spectral/adversarial losses.
They often excel at high-fidelity reconstruction and preserve prosodic cues.
EnCodec~\cite{encodec} and DAC~\cite{dac} are canonical convolutional codecs with RVQ, which have strong perceptual quality and real-time operation, but are not designed for text alignment.
Then, BigCodec~\cite{bigcodec} and TAAE~\cite{taae} achieve extremely high quality at extremely low bitrates using a convolution codec architecture.
However, they are still fixed-rate and not LM-friendly for speech–text joint modeling.
Therefore, Mimi~\cite{mimi} deliberately reduces token rates to be more LM-friendly, which is a useful bridge from pure reconstruction toward joint modeling.
In contrast, our method goes beyond purely acoustic reconstruction by jointly modeling speech tokens in both acoustic and semantic information.

\paragraph{Semantic-Acoustic Joint Modeling.}
To incorporate linguistic and semantic information and enable unified modeling, recent work blends semantic and acoustic representations.
SpeechTokenizer~\cite{speechtokenizer} distills semantics into upper RVQ codebooks using SSL features, yielding tokens that carry both semantic and acoustic content, but its sequences remain relatively long and not text-aligned.
DualCodec~\cite{dualcodec} uses dual branches, semantic and acoustic, to balance intelligibility and fidelity, with lower rates but still not explicitly aligned to text.
Spirit-LM~\cite{spiritlm} interleaves speech and text tokens for joint modeling; effective but depends on extra interleaving rules rather than addressing alignment at tokenization.
These methods add semantics or interleaving strategies, yet do not solve alignment at the tokenization stage, leaving training and inference complexity and potential prosody loss.
Our method differs by enforcing speech-text length alignment during tokenization, while still preserving both acoustic and semantic information.

\paragraph{Text-Aligned Speech Tokens.}
Text-aligned speech tokens use text positions as queries over speech representations so that tokenization itself aligns speech tokens to the text length.
Sequences become LM-friendly while retaining prosodic and paralinguistic cues.
Because of alignment, the token rate is about \(1/20\)--\(1/5\) of traditional tokenizers.
TASTE~\cite{taste} introduces the text-aligned paradigm, where text-query cross-attention over speech encoder features produces time-variant, text-length-matched tokens, thereby eliminating speech–text length mismatch and enabling straightforward SLM training.
However, the extreme compression of the bitrate and token rate results in the loss of acoustic details.
Unlike prior text-aligned approaches, our method maintains the extremely low frame rate while preserving richer acoustic detail in reconstruction.
\section{Preliminaries}

\begin{figure*}
    \centering
    \includegraphics[width=1\linewidth]{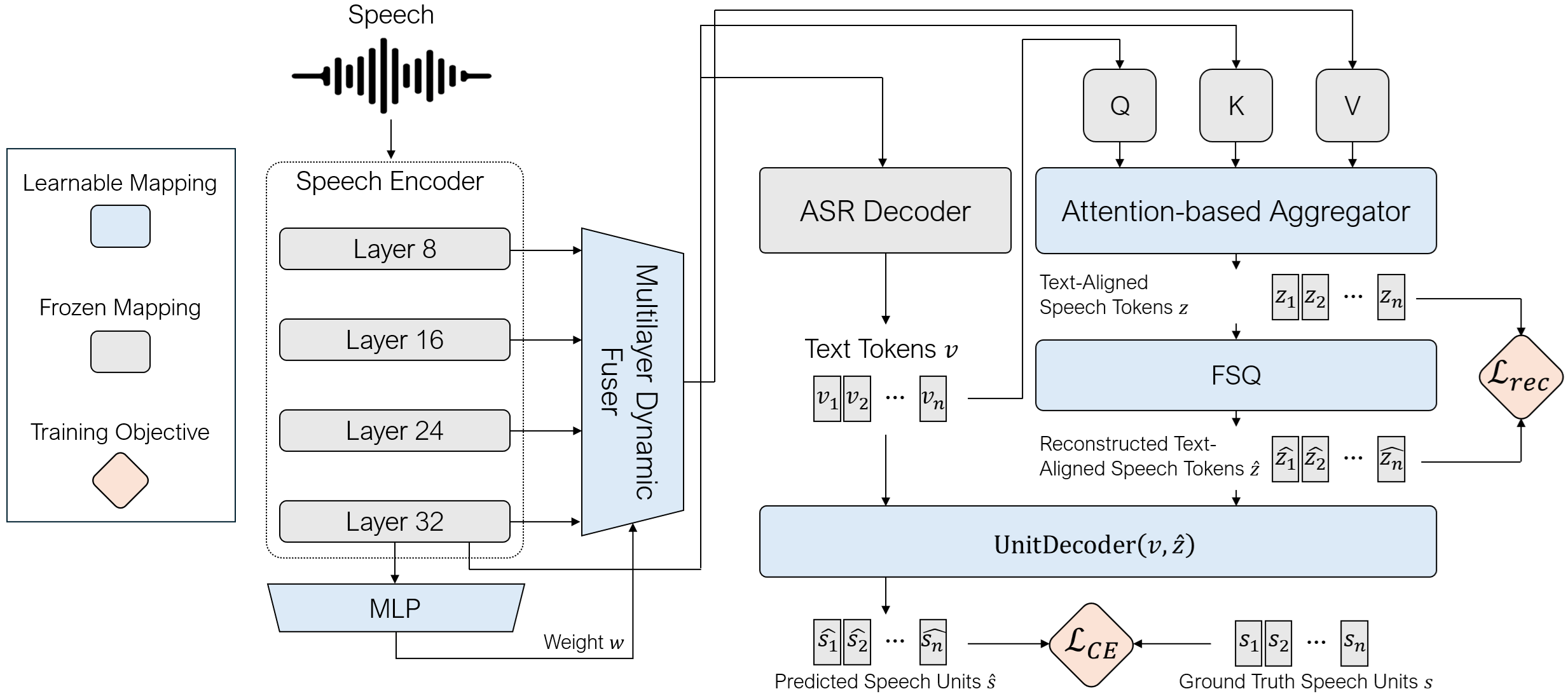}
    \caption{\textbf{TASLA overview.} Text tokens (\(Q\)) cross‑attend to the last‑layer encoder states (\(K\)) and to a dynamically mixed value stream (\(V\)) formed by per‑frame weights over layers 8/16/24/32 (MLDA). The aggregated features (\(z\)) are discretized by FSQ (\(d=64\), \(L=8\)) and fed to a unit decoder that predicts S3 units; a vocoder reconstructs waveforms. Training minimizes next‑unit cross‑entropy (\(\mathcal{L}_{CE}\)) and an FSQ reconstruction loss \((\mathcal{L}_{rec})\).}
    \label{fig:tasla}
\end{figure*}

\subsection{Speech–Text Gap}
Speech and text provide two complementary but structurally different views of language.
Let a spoken utterance be denoted as a continuous waveform \(u \in \mathbb{R}^{T}\), where \(T\) is the number of acoustic frames.
On the other hand, let the corresponding transcript be a discrete token sequence \(v = [v_1, \dots, v_N]\) of length \(N\).
The asymmetry between \(|T|\) and \(|N|\) is striking: while text unfolds sparsely with clear boundaries, speech evolves densely, carrying layered information including acoustic details and prosody.
This discrepancy yields the \emph{length mismatch problem} that most speech tokenizations produce sequences far longer than text, making it hard to jointly model text and speech.
Prior remedies, such as interleaving speech and text tokens or introducing alignment heuristics, reduce the mismatch but do not directly resolve the modality gap, and often sacrifice prosodic fidelity.

\subsection{Text-Aligned Speech Tokens}
To address this, TASTE~\cite{taste} proposes to construct speech tokens that are \emph{explicitly aligned} with their text counterpart, which is called text-aligned speech tokens.
Formally, in text-aligned speech token architectures, each time a text token is generated, a corresponding speech token is also generated, and the two token sequences are identical in length and order.
In practice, TASTE uses distilled Whisper-large-v3's~\cite{distilledWhisper} encoder part as its encoder, which is a 32-layer speech encoder.
Then, given the encoder's hidden states, \(\{h^{(1)}, \dots, h^{(L)}\}\) where each \(h^{(\ell)} \in \mathbb{R}^{T \times d_h}\), text-aligned speech tokens aim to produce a compressed representation \(z \in \mathbb{R}^{N \times d_z}\) whose length matches that of the text sequence.
A cross-attention mechanism achieves this by letting the text embeddings \(E(v) \in \mathbb{R}^{N \times d_q}\) act as queries, the final-layer states \(h^{(L)}\) provide keys, and selected shallow layer \(h^{(\ell)}\) provide values.
The aggregated representation can thus be written as
\[
z = \text{Attn}(E(v),\, h^{(L)},\, h^{(\ell)}),
\]
This formulation aligns speech and text sequences at the token level, enabling straightforward joint modeling without heuristic synchronization.

After generating text tokens \(v\) and speech tokens \(\hat{z}\), we sum them and feed the result into a speech unit decoder, which autoregressively predicts the S3 units \(\hat s\) (see \mysecref{ap:s3-units} for details).
These units are then passed to the CosyVoice unit-to-speech vocoder for waveform reconstruction.

\subsection{Quantization for Efficiency}
Even with alignment, the representation \(z\) remains continuous and high-dimensional, which hinders efficiency and symbolic control.
Discretization is therefore essential, which maps each vector \(z_i\) into a compact set of discrete codes, allowing symbolic modeling akin to text tokens.
Traditional approaches such as Residual Vector Quantization (RVQ) express each token as a sum of codebook vectors,
\[
z_i \approx \sum_{r=1}^R q^{(r)}_i, \quad q^{(r)}_i \in \mathcal{C}_r,
\]
where \(R\) is the number of quantizer stages and \(\mathcal{C}_r\) is the \(r\)-th codebook.
Although RVQ provides fine-grained reconstruction, it often requires large codebooks and multiple stages, which inflate bitrate and introduce redundancy.
This motivates alternative quantization strategies that achieve comparable expressiveness with fewer parameters, lower bitrate, and better preservation of prosodic cues.

\subsection{Bitrate and Token Rate}
\label{subsec:bitrate_tokenrate}
We distinguish two quantities: the \emph{frame rate} $R_{\text{tok}}$ (Hz) and the \emph{bitrate} $b$ (bits/s).
The token rate measures how many discrete speech tokens are produced per second for reconstruction, while the bitrate measures how many bits per second are actually needed to encode those tokens.

\paragraph{RVQ-based codecs.}
For residual vector quantization (RVQ) with $R$ quantizers and a codebook size $K$ per layer, each token consumes \(R \log_2 K,\) and the bitrate is \(\textstyle b = R_{\text{tok}} \cdot R \log_2 K.\)
Intuitively, increasing either the token frequency $R_{\text{tok}}$ or the per-token code payload $R\log_2 K$ raises the bitrate.
This RVQ counting is standard in neural codecs/speech tokenizers that output layered code indices.

\paragraph{FSQ-based tokenizers.}
For finite scalar quantization (FSQ) with $d$ scalar dimensions and $L$ uniform levels per dimension, each token carries \(d \log_2 L \quad \text{bits}\),and therefore \(\textstyle b = R_{\text{tok}} \cdot d \log_2 L\).
Compared to RVQ, FSQ trades the number of residual codebooks for axis-aligned scalar bins.
Its bitrate still scales linearly in $R_{\text{tok}}$ but now with the per-token payload $d\log_2 L$.

\paragraph{Bitrate and Frame Rate Calculation for Text-Aligned Speech Tokens.}
Since the text-aligned speech token sequence length is aligned to the text token length, it is not a fixed token rate tokenizer.
Therefore, we use the \librispeech{} test set to estimate the frame rate and calculate the bitrate.
In the \librispeech{} test set, there are about 19,805.2 seconds and a total of 51,903 text tokens; therefore, the average frame rate for text-aligned speech tokens is about 2.62.

\section{Methodology}

Our goal is to design a text-aligned speech tokenization framework that captures richer acoustic and prosodic cues while remaining simple to train and easy to control. Concretely, we (i) aggregate multi-layer encoder features with \emph{Multi-Layer Dynamic Attention} (MLDA) to compress frame-level speech into text-length representations and (ii) discretize these representations by \emph{Finite Scalar Quantization} (FSQ) to control bitrate with minimal parameters.
Finally, we train the whole system with a lightweight objective that combines a unit-level cross-entropy and a masked reconstruction loss.

\subsection{Multi-Layer Dynamic Attention}
\label{subsec:mlda}
We turn frame-level speech features into a text-length sequence by letting text tokens query the speech encoder’s hidden states.
The keys come from the last encoder layer, while the values come from shallower layers.
An MLP produces per-layer mixture weights so the model can adapt each layer's mix ratio for each frame.

We first pass the input speech \(u\) through a frozen speech encoder to obtain layerwise hidden states \(\{h^{(1)}, \dots, h^{(L)}\}\), where each \(h^{(\ell)} \in \mathbb{R}^{T \times d_h}\) is a frame-level sequence, \(T\) is acoustic length, and \(d_h\) is hidden size.

To align speech with text, we employ a cross-attention mechanism that compresses the speech frames into the same length as the text tokens. Specifically, the text tokens serve as queries \(Q\), while the last-layer representation \(h^{(L)}\) provides the keys \(K\). For the values, we select one or more shallow layers \(\{h^{(\ell_s)}\}_{s=1}^S\), denoted as \(\{V_l\}\), where each \(V_l \in \mathbb{R}^{T \times d_v}\).

To dynamically aggregate these value sources, we introduce an MLP that predicts layer-wise mixture weights. Given the last hidden state \(\mathbf{h}_{\text{last}}\), the MLP produces normalized weights as:
\[
\mathbf{w} = \mathrm{softmax}\!\left(\mathrm{MLP}\!\left(\mathbf{h}_{\text{last}}\right)\right), \quad
\mathbf{w} \in \mathbb{R}^{|\mathcal{V}|\times T}.
\]

After obtaining the weights, we use a multilayer dynamic fuser to fuse these value sources.
The aggregated value sequence is then computed frame-wise as:
\[
\tilde{V}_i = \sum_{l \in \mathcal{V}} w_{l,i} \, V_{l,i}, \quad 
\tilde{V} \in \mathbb{R}^{T \times d_v},
\]
where \(\mathcal{V}\) denotes the set of selected layers and \(i\) is the frame index.

Let \(E(\cdot)\) be a token embedding function. We embed the text sequence \(v\) into queries \(Q = E(v) \in \mathbb{R}^{N \times d_q}\), where \(N\) is the text length and \(d_q\) is the query dimension.
Finally, we compute the cross-attention between text tokens and aggregated speech tokens represented as:
\[
z=\mathrm{Attn}(Q, K, \tilde{V}) = \mathrm{softmax}\!\left(\frac{Q K^\top}{\sqrt{d_k}}\right)\tilde{V},
\]
where \(K \in \mathbb{R}^{T \times d_k}\) and \(z \in \mathbb{R}^{N \times d_v}\).

\subsection{Finite Scalar Quantization (FSQ)}
\label{subsec:fsq}
Given a per-token vector \(x\in\mathbb{R}^{D}\), we map it to a \(d\)-dim latent (via a linear encoder) and back to \(\mathbb{R}^{D}\) after quantization (via a linear decoder). FSQ operates independently on each of the \(d\) latent dimensions as follows. First, apply a learnable per-dimension affine and temperature-scaled squashing:
\[
\tilde u \;=\; u \odot s + b,\qquad
\bar u \;=\; \tanh\!\Big(\frac{\tilde u}{\tau}\Big)\ \in[-1,1]^d.
\]
With \(L\) uniform scalar levels on \([-1,1]\), \(\mathcal{G}=\{-1+\tfrac{2k}{L-1}\}_{k=0}^{L-1}\), the quantization index and value for each dimension \(j\) are
\[
i_j=\mathrm{clip}\!\left(\mathrm{round}\!\left(\frac{\bar u_j+1}{2}(L-1)\right),\,0,\,L-1\right)
\]
\[
q_j=-1+\frac{2\,i_j}{L-1}
\]
We use the straight-through estimator to form the quantized latent
\[
z_q \;=\; \bar u + (q-\bar u)_{\mathrm{sg}},
\]
where \((\cdot)_{\mathrm{sg}}\) stops gradients. The decoder maps \(z_q\) back to the feature space; training minimizes MSE on valid frames (with masking if sequences are padded).
In our experiments, we select the \(d=64\) and \(L=8\) for a balance of bitrate and performance.

\subsection{Training Objective}
\label{subsec:training_obj}
\paragraph{Cross Entropy Loss.}
After we get the text token \(v\) and speech token \(\hat z\), we uses a transformer-based unit decoder to autoregressively decode the text and speech token to speech units and then use the CosyVoice speech vocoder to convert it into speech.
Therefore, we calculate the cross-entropy loss of the target speech units.
When a target speech unit sequence $y$ is available, a unit decoder consumes $z_q$ and predicts $y$ autoregressively with the usual next-token cross-entropy:
\[
\mathcal L_{\text{CE}}
=\frac{1}{|y|}\sum_{t}-\log p_\theta\big(y_t\,\big|\,y_{<t},z_q\big).
\]

\paragraph{Reconstruction Loss.}
To further help the FSQ module learn how to reconstruct tokens back to latent vectors, we use a reconstruction loss to further help the training process.
Independent of speech units, we mask the padded frames and calculate the MSE loss of the pre-FSQ feature $z$ and its dequantized feature $z_q$:
\[
\mathcal L_{\text{recon}}
=\tfrac{1}{\sum m}\big\|(z - z_q)\odot m\big\|_2^2,
\]
where $m$ is a binary mask over valid positions/tokens.

\paragraph{Combining the Two Objectives.}
We optimize a simple weighted sum
\[
\mathcal L_{\text{total}}
\;=\; \mathcal L_{\text{CE}} \;+\; \lambda\,\mathcal L_{\text{recon}},
\]
with $\lambda$ controlling the strength of the FSQ reconstruction term.
\section{Experimental Setup}
\begin{table*}[t]
\centering
\footnotesize
\setlength{\tabcolsep}{4pt}
\renewcommand{\arraystretch}{1.1}
\begin{tabularx}{\linewidth}{l|cYYYYYY}
\toprule
\cmidrule(lr){2-8}
Model & Ene. RMSE $\downarrow$ & F0-PCC $\uparrow$ & Phr. Cos. $\uparrow$ & Ene. PCC $\uparrow$ & Phr. L2 $\downarrow$ & VDE $\downarrow$ & GPE $\downarrow$ \\
\midrule
\multicolumn{8}{l}{\librispeech{} (in-domain)}\\
\rowcolor{red!10} S3 Topline          & 6.94 & 0.91 & 0.92 & 0.95 & 5.41 & 0.15 & 0.03 \\
Text-only Baseline & 10.08 & 0.33 & 0.89 & 0.81 & 7.06 & 0.25 & 0.32 \\
TASTE               & 8.63 & 0.80 & 0.91 & 0.88 & 5.79 & 0.19 & 0.08 \\
TASLA               & \textbf{6.97} & \textbf{0.87} & 0.90 & \textbf{0.92} & 6.67 & \textbf{0.17} & \textbf{0.05} \\
\specialrule{1.2pt}{2pt}{2pt}
\multicolumn{8}{l}{\voxceleb{} (noisy/natural, OOD)}\\
\rowcolor{red!10} S3 Topline          & 5.31 & 0.86 & 0.94 & 0.94 & 4.46 & 0.17 & 0.03 \\
Text-only Baseline & 8.73 & 0.19 & 0.90 & 0.57 & 6.48 & 0.33 & 0.35 \\
TASTE               & 7.68 & 0.64 & 0.93 & 0.74 & 5.16 & 0.26 & 0.13 \\
TASLA               & \textbf{6.53} & \textbf{0.71} & \textbf{0.93} & \textbf{0.81} & \textbf{4.92} & \textbf{0.22} & \textbf{0.09} \\
\specialrule{1.2pt}{2pt}{2pt}
\multicolumn{8}{l}{\expresso{} (emotion-rich, OOD)}\\
\rowcolor{red!10} S3 Topline          & 8.57 & 0.91 & 0.82 & 0.93 & 7.19 & 0.13 & 0.04 \\
Text-only Baseline & 8.95 & 0.21 & 0.68 & 0.80 & 11.77 & 0.27 & 0.45 \\
TASTE               & 8.90 & 0.73 & 0.76 & 0.86 & 8.73 & 0.20 & 0.19 \\
TASLA               & \textbf{7.87} & \textbf{0.84} & \textbf{0.82} & \textbf{0.91} & \textbf{7.43} & \textbf{0.15} & \textbf{0.10} \\
\specialrule{1.2pt}{2pt}{2pt}
\bottomrule
\end{tabularx}
\caption{\textbf{Experimental Results for Prosody Metrics.} We report prosody metrics across datasets at text-aligned speech token architectures.
Our proposed method, \method{}, outperforms the ablations and baselines on nearly all prosody metrics. The rows marked in light red denote the upper bound (S3 topline).}
\label{tab:prosody_only}
\end{table*}

\subsection{Datasets}
We conduct experiments on both in-domain and out-of-domain datasets to comprehensively evaluate our framework.
\librispeech{}~\cite{librispeech} is used for model training and primary evaluation, while \expresso{}~\cite{expresso} and \voxceleb{}~\cite{voxceleb} are adopted for out-of-domain evaluation, as they contain richer acoustic variability aligned with our objectives.
Specifically, \expresso{} emphasizes emotional expressiveness and prosodic variation, whereas \voxceleb{} features diverse recording conditions with natural background noise.
The detailed dataset introduction could be found in \mysecref{ap:dataset}

\subsection{Baselines}

Our study focuses on \emph{text-aligned} speech tokenization, aiming to produce token sequences whose length matches the text while keeping the bitrate low enough for joint speech-text modeling. To evaluate \method{}, we consider representative tokenizers from three major categories: Compression Bitrate Neural Codecs, Semantic-Acoustic Joint Modeling Tokenizers, and Text-Aligned Speech Tokenizers. 

Since many existing codecs operate at bitrates much higher than 1 kbps, we downsample their outputs—by retaining fewer RVQ layers or applying temporal striding—to approximately 1 kbps. This allows us to ensure fair comparisons under a comparable bitrate regime. The implementation details are provided in \mysecref{ap:baselines}.

Because \method{} compresses the token rate to an extremely low level while preserving text alignment, there is no directly comparable tokenizer besides TASTE. We therefore include state-of-the-art codecs that either operate at, or can be adjusted to, similar bitrates: EnCodec~\cite{encodec}, Mimi~\cite{mimi}, SpeechTokenizer~\cite{speechtokenizer}, TASTE~\cite{taste}, DualCodec~\cite{dualcodec}, and BigCodec~\cite{bigcodec}. These baselines vary in training data and parameter scales, which we summarize in \mysecref{ap:baselines}. Our conclusions are thus scoped to this targeted bitrate regime rather than claiming universal superiority.

For ablation studies, we further introduce the S3 topline and a text-only baseline. The S3 topline reconstructs speech directly from ground-truth S3 units, while the text-only baseline predicts S3 tokens from text without joint modeling.

\subsection{Evaluation Metrics}

We evaluate our framework along two complementary dimensions: \emph{quality}, which reflects the overall naturalness and fidelity of the reconstructed audio, and \emph{prosody}, which measures how well prosodic patterns such as pitch, rhythm, and energy are preserved.

\subsubsection{Prosody Metrics}
\label{subsec:prosody_metrics}
For prosody metrics, we report Emotion Consistency, F0-PCC, Energy-PCC, Energy-RMSE, Phrase Cosine Similarity (Phr. Cos.), Phrase L2 (Phr. L2), Voicing Decision Error (VDE), and Gross Pitch Error (GPE). Emotion Consistency checks whether the emotional intent is preserved after reconstruction. F0-PCC measures the correlation of pitch contours between hypothesis and reference after voicing masking and time alignment, while Energy-PCC and Energy-RMSE quantify how well loudness dynamics over time are retained. Phr. Cos. and Phr. L2 summarize phrase-level intonation by converting F0 to semitones, fitting degree-3 Legendre polynomials on normalized time, and comparing the resulting coefficient vectors of reference and reconstruction. VDE is the fraction of frames with mismatched voiced/unvoiced decisions (lower is better), and GPE is the proportion of voiced frames where predicted F0 deviates from the reference by a large relative margin, reflecting robustness of pitch accuracy.

\subsubsection{Quality Metrics}
For quality metrics, we use Word Error Rate (WER), UTMOS, and Speaker Similarity to evaluate the reconstructed speech's quality.
Word Error Rate (WER) is computed between the ASR transcription of the reconstructed audio and the reference transcript.
UTMOS~\cite{utmosv2} is a non-intrusive, neural MOS predictor that estimates perceived naturalness and quality directly from the waveform.
Speaker similarity is quantified as the cosine similarity between speaker embeddings extracted from reference and reconstructed audio.

\begin{table*}[t]
\centering
\scriptsize
\renewcommand{\arraystretch}{1.08}
\setlength{\tabcolsep}{3pt}
\begin{tabularx}{\linewidth}{l|c|Y|YYY|YY|YYY}
\toprule
& & & \multicolumn{3}{c|}{\librispeech{}} & \multicolumn{2}{c|}{\voxceleb{} (OOD)} & \multicolumn{3}{c}{\expresso{} (OOD)} \\
\cmidrule(lr){4-6} \cmidrule(lr){7-8} \cmidrule(lr){9-11}
Model & Frame Rate & Bitrate & WER & UTMOS & Spk. Sim. & UTMOS & Spk. Sim. & WER & UTMOS & Spk. Sim. \\
\midrule
Ground Truth & --- & 256000 & 0.05 & 3.41 & ---  & 2.82 & --- & 0.11 & 2.98 & --- \\
\rowcolor{gray!20}\multicolumn{11}{c}{\emph{frame rate less than 150}}\\
EnCodec    & 75 & 1500 & 0.07 & 1.40 & 0.63 & 1.44 & 0.61 & 0.13 & 1.11 & 0.56 \\
SpeechTokenizer & 50 & 1000 & 0.11 & 1.72 & 0.34 & 1.80 & 0.39 & 0.30 & 1.51 & 0.34 \\
BigCodec   & 80 & 1040 & 0.05 & 3.29 & 1.00 &  2.76 & 0.99 & 0.12 & 2.79 & 0.99 \\
DualCodec  & 12.5 & 1225 & 0.04 & 3.16 & 0.88 &  2.60 & 0.84 & 0.09 & 2.80 & 0.84 \\
Mimi       & 12.5 & 1100 & 0.04 & 3.01 & 0.93 &  2.53 & 0.93 & 0.10 & 2.45 & 0.89 \\
\rowcolor{gray!20}\multicolumn{11}{c}{\emph{frame rate less than 10 (word-level)}} \\
\mbox{S3 Topline} & --- & 600 & 0.04 & 3.40 & 0.87 &  2.93 & 0.84 & 0.12 & 3.17 & 0.82 \\
\mbox{Text-only Baseline} & --- & 50 & 0.23 & 3.42 & 0.77 &  3.33 & 0.68 & 0.41 & 3.28 & 0.64 \\
TASTE      & \(\sim\)2.62 & \(\sim\)150 & 0.10 & 3.51 & 0.85 &  3.31 & 0.78 & 0.27 & 3.33 & 0.75 \\
TASLA      & \(\sim\)2.62 & \(\sim\)600 & 0.12 & 3.43 & 0.87 & 3.10 & 0.81 & 0.28 & 3.12 & 0.80 \\
\specialrule{1.2pt}{2pt}{2pt}
\bottomrule
\end{tabularx}
\caption{\textbf{Experimental Results for Quality Metrics.} Lower is better for WER; higher is better for the others. \voxceleb{} has no WER column because no transcripts are available. Our proposed method (\method{}) outperforms or performs competitively against the baselines over different evaluation sets.
The out-of-domain (OOD) tag is applied to TASTE and TASLA in this table, since the two OOD datasets may overlap with the training data of other speech tokenizers.
}
\label{tab:quality}
\end{table*}

\section{Experimental Results}

We report the main experimental results of the quality metrics in \mytable{tab:quality} and prosody metrics in \mytable{tab:prosody_only}.
The details of the weight analysis could be found in \mysecref{subsec:dynamic_weight}.
The details of the training and parameters can be found in \mysecref{ap:training_details}.
The details of metrics formulation and explanation could be found in \mysecref{ap:metrics}.

\subsection{Overall Performance}
From \mytable{tab:prosody_only}, \method{} consistently surpasses other text-aligned speech token frameworks on nearly every prosody indicator.
And it is also closer to the S3 tokens' topline.
This pattern shows that \method{} preserves paralinguistic cues, intonation, rhythm, and emphasis, rather than merely reconstructing clean audio, and does so under much stronger compression.
For pitch-related behavior, a higher F0-PCC means \method{}’s generated pitch contour follows the reference more faithfully.
At the same time, lower GPE and VDE indicate fewer octave (pitch halving/doubling) mistakes and more stable voiced/unvoiced decisions, yielding more natural intonation and fewer pitch glitches.
For loudness dynamics and phrasing, higher Energy-PCC reflects a closer alignment of energy fluctuations over time, while lower Energy-RMSE shows that the absolute loudness levels better match the reference, rather than just the trend.
Higher Phrase-Cos. and lower Phrase-L2 further indicate that pauses, timing, and rhythmic structure align more precisely with the ground truth.

On the other hand, \mytable{tab:quality} summarizes the quality results across in-domain \librispeech{} and two out-of-domain datasets, \voxceleb{} and \expresso{}. 
Overall, \method{} achieves superior performance compared to other commonly used speech codecs under most of the quality metrics under a significantly low frame rate.
Compared to other low frame rate methods, \method{} is also closer to the S3 topline.
Specifically, on the OOD sets, the higher frame rate speech codecs get more advantage on WER since they could reconstruct more acoustic details, and the training data of the ASR models mostly contains \librispeech{} and a large amount of training data.
Therefore, even though the quality of the reconstructed speech of other methods is lower than \method{}, they could also get lower WER on these evaluation sets.

These improvements match our design goal: text-aligned speech tokens that retain prosody while remaining LM-friendly under aggressive compression.
Mechanistic analysis for this behavior is provided in \mysecref{subsec:dynamic_weight}.

\subsection{Analysis of Dynamic Weights}
\label{subsec:dynamic_weight}

\begin{figure*}
    \centering
    \includegraphics[width=1\linewidth]{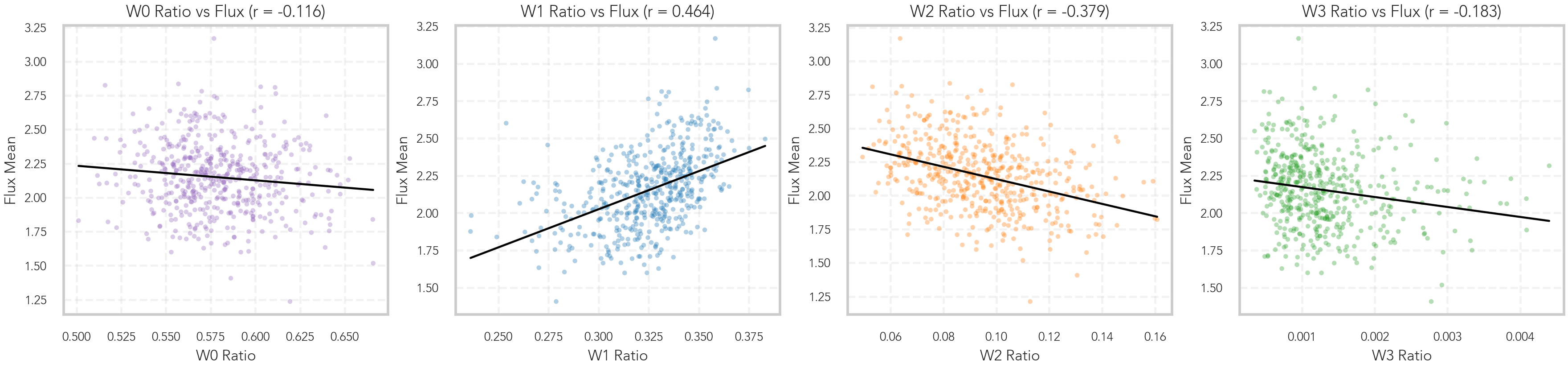}
    \caption{\textbf{Scatter plots and linear regression across all samples.} The figure shows that \(w_1\) is positively correlated with mean spectral flux, while \(w_2\) is negatively correlated. This indicates that \(w_2\) suppresses spectral transients, whereas \(w_1\) promotes them.}
    \label{fig:flux}
\end{figure*}

\begin{figure}
    \centering
    \includegraphics[width=1\linewidth]{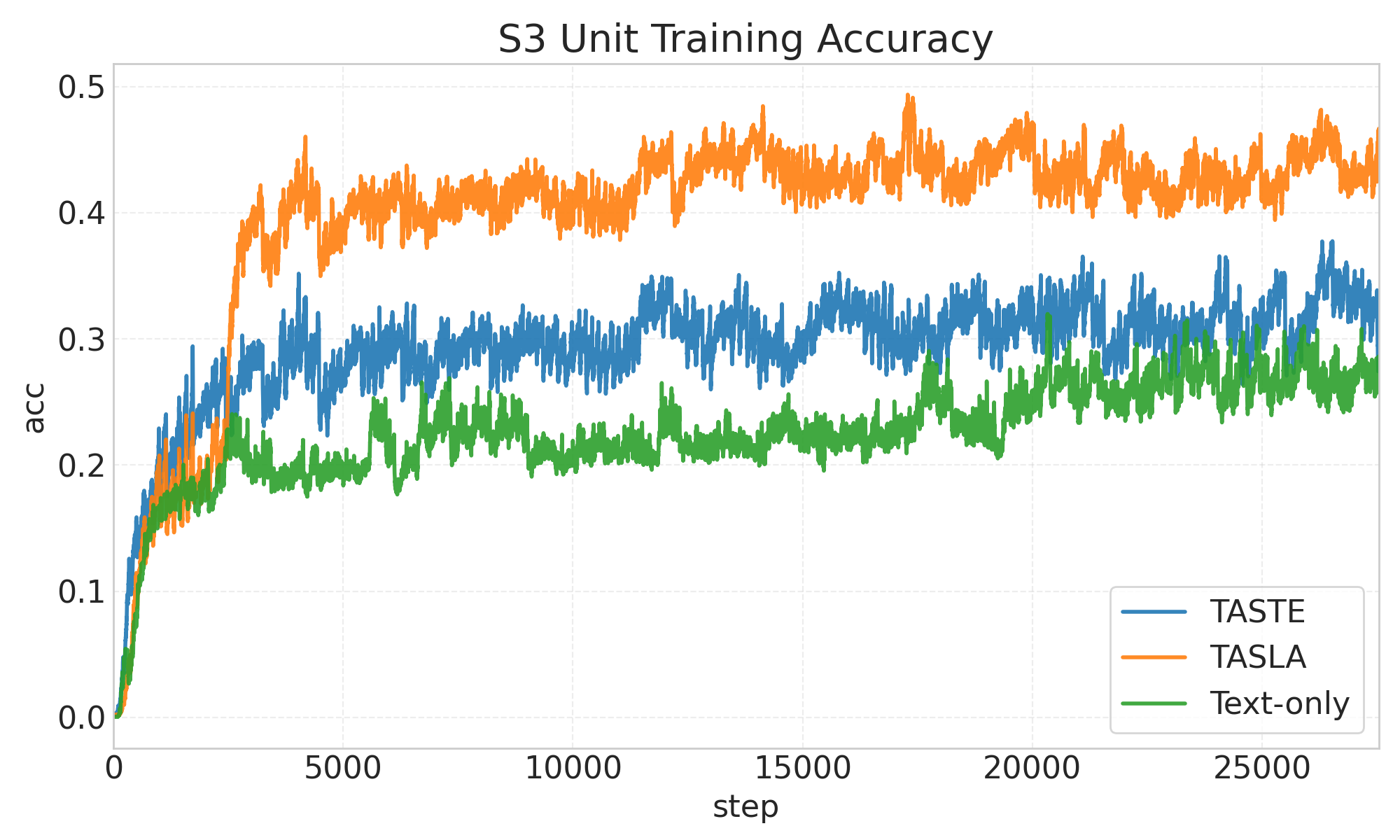}
    \caption{\textbf{S3 Unit Training Accuracy.} Accuracy of S3 unit prediction for the three ablations: Text-only, TASTE, and TASLA. TASLA achieves about 10\% higher accuracy than Text-only and TASTE.
}
    \label{fig:s3unit}
\end{figure}

In our experiments, we select the 8th, 16th, 24th, and 32nd layers as the target layers for multilayer dynamic attention.
In this section, we use \(\{w_0, w_1, w_2, w_3\}\) to represent these weights.

We analyze 500 utterances from \expresso{} to characterize the dynamic weighting module, since \expresso{} is richer with prosody.
The per-frame normalized weight means are \(\bar{w}_0=0.573,\ \bar{w}_1=0.325,\ \bar{w}_2=0.100,\ \bar{w}_3=0.0017\).
Thus the stream is carried predominantly by \(w_0\), with \(w_1\) and \(w_2\) providing substantial secondary contribution.
To quantify how many layers act at once, we compute frame-wise entropy
\[
H_t=-\sum_{i=1}^{4} w_t^{(i)}\log w_t^{(i)}
\]
and report \(\bar H=0.484\).
The effective number of layers \(\mathrm{ENL}=\exp(\bar H){\approx}1.62\) indicates about one to two meaningful contributors per frame.

We relate weights to short-time spectral dynamics.
Let spectral flux be
\(
F_t=\lVert\mathbf m_t-\mathbf m_{t-1}\rVert_2
\), where \(\mathbf m_t\) denotes mel magnitudes.
High \(F_t\) marks onsets/offsets and low \(F_t\) marks steady vowels or near-silence.
From \myfig{fig:flux}, we observe \(\mathrm{corr}(w_0, F)=-0.116\), \(\mathrm{corr}(w_1, F)=+0.464\), \(\mathrm{corr}(w_2 ,F)=-0.379\), and \(\mathrm{corr}(w_3 ,F)=-0.183\).
Previous work~\citep{DBLP:journals/speech/BabyPSM20} denotes that the spectrum flux has a positive relation with the phoneme or sub-word boundaries. 
Combining with the observation of the average weight and the relation from \myfig{fig:s3unit}, deeper layers (16th and 24th layer, \(w_1\) and \(w_2\)) are more related to the flux; therefore, the deeper layers may relate to the phoneme or sub-word boundaries.
On the other hand, the \(w_0\) seems not obviously related to the spectrum flux, but has a high weight ratio, which shows that shallow layers provide more acoustic cues, but are not obviously related to phoneme or sub-word boundaries.
Therefore, since the semantic information has been carried with the text tokens, the deeper layer's weight becomes smaller since it carries lower acoustic cues.
This observation also matches the conclusion of~\citet{DBLP:conf/naacl/MaQFTGK25} and~\citet{DBLP:journals/corr/abs-2505-19606}.
Therefore, we provide an ablation study that mainly uses shallower layers for MLDA; see the experiments in~\mysecref{ap:shallow}.

\subsection{S3 Unit Accuracy}
From \myfig{fig:s3unit}, we observe that \method{} predicts the S3 units more accurately than TASTE and text-only ablations.
These results indicate that \method{} is better at jointly modeling acoustic and semantic features, leading to more accurate S3 unit predictions.
However, this improvement is not directly reflected in the quality metrics but instead appears in the prosody metrics.
This limitation may be due to the upper bound of the S3 units’ performance and the size of the training data.

Combining these observations, the combination of the observed weight switch pattern with flux and the S3 prediction accuracy, provides a mechanistic explanation for the dynamically fused speech features and their importance.
\section{Conclusion}
We introduce TASLA, a text-aligned speech tokenization framework designed to preserve fine-grained acoustic detail under an extremely low frame rate while remaining compatible with text-token alignment.
Our experiments demonstrate that, despite the time-variant nature of the speech tokens, TASLA consistently maintains rich acoustic cues, achieving prosody metrics close to the upper-bound topline and outperforming prior text-aligned speech tokenizers on out-of-domain tasks. Quantitative analyses reveal how different encoder layers contribute to various acoustic aspects, offering interpretability that helps explain why certain layers dominate under specific speaking conditions.
Future work includes improving time efficiency through low-latency variants, and exploring single-stage generation approaches to overcome the performance ceiling imposed by S3 speech units.
\section{Limitations}
Due to the performance ceiling of the S3 units and the limited dataset size, \method{} has untapped potential to model speech more effectively than other text-aligned speech tokenizers, achieving higher accuracy in predicting S3 units. While this advantage does not manifest directly in quality metrics, it is reflected in improved prosody performance and a narrower gap to the S3 units’ topline.

\paragraph{Ethical Considerations and Potential Risks.}
While TASLA focuses on text-aligned speech tokenization rather than speech generation, discrete speech representations could be misused for voice spoofing or cloning. All models and data used in this study are for research purposes only, and all datasets are public and released under academic licenses. We encourage responsible use and adherence to ethical standards when reproducing or extending this work.
We used AI assistants for grammar polishing and document formatting only. All experimental design, implementation, and data analysis were manually verified by the authors.

\bibliography{custom}

\begin{thebibliography}{32}
\providecommand{\natexlab}[1]{#1}

\bibitem[{Arora et~al.(2025)Arora, Chang, Chien, Peng, Wu, Adi, Dupoux, Lee, Livescu, and Watanabe}]{DBLP:journals/corr/abs-2504-08528}
Siddhant Arora, Kai{-}Wei Chang, Chung{-}Ming Chien, Yifan Peng, Haibin Wu, Yossi Adi, Emmanuel Dupoux, Hung{-}Yi Lee, Karen Livescu, and Shinji Watanabe. 2025.
\newblock On the landscape of spoken language models: {A} comprehensive survey.
\newblock \emph{CoRR}, abs/2504.08528.

\bibitem[{Baba et~al.(2024)Baba, Nakata, Saito, and Saruwatari}]{utmosv2}
Kaito Baba, Wataru Nakata, Yuki Saito, and Hiroshi Saruwatari. 2024.
\newblock The {T05} system for the voicemos challenge 2024: Transfer learning from deep image classifier to naturalness {MOS} prediction of high-quality synthetic speech.
\newblock In \emph{{SLT}}, pages 818--824. {IEEE}.

\bibitem[{Baby et~al.(2020)Baby, Prakash, Subramanian, and Murthy}]{DBLP:journals/speech/BabyPSM20}
Arun Baby, Jeena~J. Prakash, Aswin~Shanmugam Subramanian, and Hema~A. Murthy. 2020.
\newblock Significance of spectral cues in automatic speech segmentation for indian language speech synthesizers.
\newblock \emph{Speech Commun.}, 123:10--25.

\bibitem[{Borsos et~al.(2023)Borsos, Marinier, Vincent, Kharitonov, Pietquin, Sharifi, Roblek, Teboul, Grangier, Tagliasacchi, and Zeghidour}]{audiolm}
Zal{\'{a}}n Borsos, Rapha{\"{e}}l Marinier, Damien Vincent, Eugene Kharitonov, Olivier Pietquin, Matthew Sharifi, Dominik Roblek, Olivier Teboul, David Grangier, Marco Tagliasacchi, and Neil Zeghidour. 2023.
\newblock Audiolm: {A} language modeling approach to audio generation.
\newblock \emph{{IEEE} {ACM} Trans. Audio Speech Lang. Process.}, 31:2523--2533.

\bibitem[{Chen et~al.(2024)Chen, Liu, Zhou, Liu, Tan, Li, Zhao, Qian, and Wei}]{wang2023neuralcodeclanguagemodels}
Sanyuan Chen, Shujie Liu, Long Zhou, Yanqing Liu, Xu~Tan, Jinyu Li, Sheng Zhao, Yao Qian, and Furu Wei. 2024.
\newblock {VALL-E} 2: Neural codec language models are human parity zero-shot text to speech synthesizers.
\newblock \emph{CoRR}, abs/2406.05370.

\bibitem[{D{\'{e}}fossez et~al.(2023)D{\'{e}}fossez, Copet, Synnaeve, and Adi}]{encodec}
Alexandre D{\'{e}}fossez, Jade Copet, Gabriel Synnaeve, and Yossi Adi. 2023.
\newblock High fidelity neural audio compression.
\newblock \emph{Trans. Mach. Learn. Res.}, 2023.

\bibitem[{D{\'{e}}fossez et~al.(2024)D{\'{e}}fossez, Mazar{\'{e}}, Orsini, Royer, P{\'{e}}rez, J{\'{e}}gou, Grave, and Zeghidour}]{mimi}
Alexandre D{\'{e}}fossez, Laurent Mazar{\'{e}}, Manu Orsini, Am{\'{e}}lie Royer, Patrick P{\'{e}}rez, Herv{\'{e}} J{\'{e}}gou, Edouard Grave, and Neil Zeghidour. 2024.
\newblock Moshi: a speech-text foundation model for real-time dialogue.
\newblock \emph{CoRR}, abs/2410.00037.

\bibitem[{Du et~al.(2024)Du, Chen, Zhang, Hu, Lu, Yang, Hu, Zheng, Gu, Ma, Gao, and Yan}]{cosyvoice}
Zhihao Du, Qian Chen, Shiliang Zhang, Kai Hu, Heng Lu, Yexin Yang, Hangrui Hu, Siqi Zheng, Yue Gu, Ziyang Ma, Zhifu Gao, and Zhijie Yan. 2024.
\newblock Cosyvoice: {A} scalable multilingual zero-shot text-to-speech synthesizer based on supervised semantic tokens.
\newblock \emph{CoRR}, abs/2407.05407.

\bibitem[{Gandhi et~al.(2023)Gandhi, von Platen, and Rush}]{distilledWhisper}
Sanchit Gandhi, Patrick von Platen, and Alexander~M. Rush. 2023.
\newblock Distil-whisper: Robust knowledge distillation via large-scale pseudo labelling.
\newblock \emph{CoRR}, abs/2311.00430.

\bibitem[{Guo et~al.(2025)Guo, Li, Wang, Li, Shao, Zhang, Du, Chen, Liu, and Yu}]{DBLP:journals/corr/abs-2502-06490}
Yiwei Guo, Zhihan Li, Hankun Wang, Bohan Li, Chongtian Shao, Hanglei Zhang, Chenpeng Du, Xie Chen, Shujie Liu, and Kai Yu. 2025.
\newblock Recent advances in discrete speech tokens: {A} review.
\newblock \emph{CoRR}, abs/2502.06490.

\bibitem[{Hassid et~al.(2023)Hassid, Remez, Nguyen, Gat, Conneau, Kreuk, Copet, D{\'{e}}fossez, Synnaeve, Dupoux, Schwartz, and Adi}]{hassid2024textuallypretrainedspeechlanguage}
Michael Hassid, Tal Remez, Tu~Anh Nguyen, Itai Gat, Alexis Conneau, Felix Kreuk, Jade Copet, Alexandre D{\'{e}}fossez, Gabriel Synnaeve, Emmanuel Dupoux, Roy Schwartz, and Yossi Adi. 2023.
\newblock Textually pretrained speech language models.
\newblock In \emph{NeurIPS}.

\bibitem[{Kim et~al.(2018)Kim, Salamon, Li, and Bello}]{crepe}
Jong~Wook Kim, Justin Salamon, Peter Li, and Juan~Pablo Bello. 2018.
\newblock Crepe: {A} convolutional representation for pitch estimation.
\newblock In \emph{{ICASSP}}, pages 161--165. {IEEE}.

\bibitem[{KimiTeam et~al.(2025)KimiTeam, Ding, Ju, Leng, Liu, Liu, Shang, Shen, Song, Tan, Tang, Wang, Wei, Xin, Xu, Yu, Zhang, Zhou, Charles, Chen, Chen, Du, He, Hu, Lai, Li, Liu, Sun, Wang, Wang, Wu, Wu, Yang, Yang, Yang, Yang, Yin, Yuan, Zhang, and Zhou}]{kimiteam2025kimiaudiotechnicalreport}
KimiTeam, Ding Ding, Zeqian Ju, Yichong Leng, Songxiang Liu, Tong Liu, Zeyu Shang, Kai Shen, Wei Song, Xu~Tan, Heyi Tang, Zhengtao Wang, Chu Wei, Yifei Xin, Xinran Xu, Jianwei Yu, Yutao Zhang, Xinyu Zhou, Y.~Charles, and 21 others. 2025.
\newblock Kimi-audio technical report.
\newblock \emph{CoRR}, abs/2504.18425.

\bibitem[{Kumar et~al.(2023)Kumar, Seetharaman, Luebs, Kumar, and Kumar}]{dac}
Rithesh Kumar, Prem Seetharaman, Alejandro Luebs, Ishaan Kumar, and Kundan Kumar. 2023.
\newblock High-fidelity audio compression with improved {RVQGAN}.
\newblock In \emph{NeurIPS}.

\bibitem[{Li et~al.(2025)Li, Lin, Li, Huang, Wang, Wang, Zhan, and Wu}]{dualcodec}
Jiaqi Li, Xiaolong Lin, Zhekai Li, Shixi Huang, Yuancheng Wang, Chaoren Wang, Zhenpeng Zhan, and Zhizheng Wu. 2025.
\newblock Dualcodec: {A} low-frame-rate, semantically-enhanced neural audio codec for speech generation.
\newblock \emph{CoRR}, abs/2505.13000.

\bibitem[{Liu et~al.(2024)Liu, Xu, Yuan, Wu, Wang, and Plumbley}]{DBLP:journals/jstsp/LiuXYWWP24}
Haohe Liu, Xuenan Xu, Yi~Yuan, Mengyue Wu, Wenwu Wang, and Mark~D. Plumbley. 2024.
\newblock Semanticodec: An ultra low bitrate semantic audio codec for general sound.
\newblock \emph{{IEEE} J. Sel. Top. Signal Process.}, 18(8):1448--1461.

\bibitem[{Ma et~al.(2025)Ma, Qian, Fathullah, Tang, Gales, and Knill}]{DBLP:conf/naacl/MaQFTGK25}
Rao Ma, Mengjie Qian, Yassir Fathullah, Siyuan Tang, Mark J.~F. Gales, and Kate~M. Knill. 2025.
\newblock Cross-lingual transfer learning for speech translation.
\newblock In \emph{{NAACL} (Short Papers)}, pages 33--43. Association for Computational Linguistics.

\bibitem[{Mentzer et~al.(2024)Mentzer, Minnen, Agustsson, and Tschannen}]{fsq}
Fabian Mentzer, David Minnen, Eirikur Agustsson, and Michael Tschannen. 2024.
\newblock Finite scalar quantization: {VQ-VAE} made simple.
\newblock In \emph{{ICLR}}. OpenReview.net.

\bibitem[{Nagrani et~al.(2017)Nagrani, Chung, and Zisserman}]{voxceleb}
Arsha Nagrani, Joon~Son Chung, and Andrew Zisserman. 2017.
\newblock Voxceleb: {A} large-scale speaker identification dataset.
\newblock In \emph{{INTERSPEECH}}, pages 2616--2620. {ISCA}.

\bibitem[{Nguyen et~al.(2023)Nguyen, Hsu, D'Avirro, Shi, Gat, Fazel{-}Zarandi, Remez, Copet, Synnaeve, Hassid, Kreuk, Adi, and Dupoux}]{expresso}
Tu~Anh Nguyen, Wei{-}Ning Hsu, Antony D'Avirro, Bowen Shi, Itai Gat, Maryam Fazel{-}Zarandi, Tal Remez, Jade Copet, Gabriel Synnaeve, Michael Hassid, Felix Kreuk, Yossi Adi, and Emmanuel Dupoux. 2023.
\newblock Expresso: {A} benchmark and analysis of discrete expressive speech resynthesis.
\newblock In \emph{{INTERSPEECH}}, pages 4823--4827. {ISCA}.

\bibitem[{Nguyen et~al.(2024{\natexlab{a}})Nguyen, Muller, Yu, Costa{-}juss{\`{a}}, Elbayad, Popuri, Duquenne, Algayres, Mavlyutov, Gat, Synnaeve, Pino, Sagot, and Dupoux}]{nguyen2024spiritlminterleavedspoken}
Tu~Anh Nguyen, Benjamin Muller, Bokai Yu, Marta~R. Costa{-}juss{\`{a}}, Maha Elbayad, Sravya Popuri, Paul{-}Ambroise Duquenne, Robin Algayres, Ruslan Mavlyutov, Itai Gat, Gabriel Synnaeve, Juan Pino, Beno{\^{\i}}t Sagot, and Emmanuel Dupoux. 2024{\natexlab{a}}.
\newblock Spirit-lm: Interleaved spoken and written language model.
\newblock \emph{CoRR}, abs/2402.05755.

\bibitem[{Nguyen et~al.(2024{\natexlab{b}})Nguyen, Muller, Yu, Costa{-}juss{\`{a}}, Elbayad, Popuri, Duquenne, Algayres, Mavlyutov, Gat, Synnaeve, Pino, Sagot, and Dupoux}]{spiritlm}
Tu~Anh Nguyen, Benjamin Muller, Bokai Yu, Marta~R. Costa{-}juss{\`{a}}, Maha Elbayad, Sravya Popuri, Paul{-}Ambroise Duquenne, Robin Algayres, Ruslan Mavlyutov, Itai Gat, Gabriel Synnaeve, Juan Pino, Beno{\^{\i}}t Sagot, and Emmanuel Dupoux. 2024{\natexlab{b}}.
\newblock Spirit-lm: Interleaved spoken and written language model.
\newblock \emph{CoRR}, abs/2402.05755.

\bibitem[{Panayotov et~al.(2015)Panayotov, Chen, Povey, and Khudanpur}]{librispeech}
Vassil Panayotov, Guoguo Chen, Daniel Povey, and Sanjeev Khudanpur. 2015.
\newblock Librispeech: An {ASR} corpus based on public domain audio books.
\newblock In \emph{{ICASSP}}, pages 5206--5210. {IEEE}.

\bibitem[{Parker et~al.(2025)Parker, Smirnov, Pons, Carr, Zukowski, Evans, and Liu}]{taae}
Julian~D. Parker, Anton Smirnov, Jordi Pons, CJ~Carr, Zack Zukowski, Zach Evans, and Xubo Liu. 2025.
\newblock Scaling transformers for low-bitrate high-quality speech coding.
\newblock In \emph{{ICLR}}. OpenReview.net.

\bibitem[{Radford et~al.(2023)Radford, Kim, Xu, Brockman, McLeavey, and Sutskever}]{whisper}
Alec Radford, Jong~Wook Kim, Tao Xu, Greg Brockman, Christine McLeavey, and Ilya Sutskever. 2023.
\newblock Robust speech recognition via large-scale weak supervision.
\newblock In \emph{{ICML}}, volume 202 of \emph{Proceedings of Machine Learning Research}, pages 28492--28518. {PMLR}.

\bibitem[{Shikhar et~al.(2025)Shikhar, Kurpath, Mullappilly, Lahoud, Khan, Anwer, Khan, and Cholakkal}]{DBLP:conf/acl/ShikharKMLKA0C25}
Sambal Shikhar, Mohammed~Irfan Kurpath, Sahal~Shaji Mullappilly, Jean Lahoud, Fahad~Shahbaz Khan, Rao~Muhammad Anwer, Salman~H. Khan, and Hisham Cholakkal. 2025.
\newblock Llmvox: Autoregressive streaming text-to-speech model for any {LLM}.
\newblock In \emph{{ACL} (Findings)}, pages 20481--20493. Association for Computational Linguistics.

\bibitem[{Shim et~al.(2025)Shim, Cristofaro, Hu, Vietti, and Plank}]{DBLP:journals/corr/abs-2505-19606}
Ryan~Soh{-}Eun Shim, Domenico~De Cristofaro, Chengzhi~Martin Hu, Alessandro Vietti, and Barbara Plank. 2025.
\newblock Languages in multilingual speech foundation models align both phonetically and semantically.
\newblock \emph{CoRR}, abs/2505.19606.

\bibitem[{Tseng et~al.(2025)Tseng, Chen, Lee, Shiu, and Lee}]{taste}
Liang{-}Hsuan Tseng, Yi{-}Chang Chen, Kuan{-}Yi Lee, Da{-}Shan Shiu, and Hung{-}yi Lee. 2025.
\newblock {TASTE:} text-aligned speech tokenization and embedding for spoken language modeling.
\newblock \emph{CoRR}, abs/2504.07053.

\bibitem[{Wang et~al.(2023)Wang, Chen, Wu, Zhang, Zhou, Liu, Chen, Liu, Wang, Li, He, Zhao, and Wei}]{valle}
Chengyi Wang, Sanyuan Chen, Yu~Wu, Ziqiang Zhang, Long Zhou, Shujie Liu, Zhuo Chen, Yanqing Liu, Huaming Wang, Jinyu Li, Lei He, Sheng Zhao, and Furu Wei. 2023.
\newblock Neural codec language models are zero-shot text to speech synthesizers.
\newblock \emph{CoRR}, abs/2301.02111.

\bibitem[{Xin et~al.(2024)Xin, Tan, Takamichi, and Saruwatari}]{bigcodec}
Detai Xin, Xu~Tan, Shinnosuke Takamichi, and Hiroshi Saruwatari. 2024.
\newblock Bigcodec: Pushing the limits of low-bitrate neural speech codec.
\newblock \emph{CoRR}, abs/2409.05377.

\bibitem[{Yazdani et~al.(2025)Yazdani, Ansari, Mahajan, Afsharrad, and Mousavi}]{DBLP:journals/corr/abs-2507-16835}
Nima Yazdani, Ali Ansari, Aruj Mahajan, Amirhossein Afsharrad, and Seyed~Shahabeddin Mousavi. 2025.
\newblock Evaluating speech-to-text x {LLM} x text-to-speech combinations for {AI} interview systems.
\newblock \emph{CoRR}, abs/2507.16835.

\bibitem[{Zhang et~al.(2023)Zhang, Zhang, Li, Zhou, and Qiu}]{speechtokenizer}
Xin Zhang, Dong Zhang, Shimin Li, Yaqian Zhou, and Xipeng Qiu. 2023.
\newblock Speechtokenizer: Unified speech tokenizer for speech large language models.
\newblock \emph{CoRR}, abs/2308.16692.

\end{thebibliography}

\newpage
\appendix
\appendix
\section*{Appendix}
\label{sec:appendix}
\section{Baselines and Implementation Details}
\label{ap:baselines}

\paragraph{Scope and selection.}
Our study targets \emph{text-aligned} speech tokenization suitable for joint speech–text modeling at low bitrates and short sequences. Because only TASTE is directly text-aligned at word length, we include it and compare against state-of-the-art neural codecs that either operate near, or can be adjusted to, a similar low-bitrate regime (via retaining fewer RVQ layers and/or applying temporal striding) to ensure a fair comparison. We also report two ablations (S3 topline and text-only). The baselines used throughout the paper are: EnCodec, Mimi, SpeechTokenizer, DualCodec, BigCodec, and TASTE, plus the S3 topline and our text-only baseline.

\paragraph{Common evaluation wrapper.}
All baselines are run under a unified path-based pipeline: mono \texttt{float32} waveforms are loaded from disk, resampled to each model’s native sampling rate, encoded into discrete codes/units, and decoded back to waveforms. For RVQ/stacked-codebook models, we probe multiple operating points by (i) keeping only the first $k$ quantizer layers and/or (ii) applying temporal sub-sampling, to bring effective rates into a comparable low-bitrate regime for joint modeling.

\paragraph{EnCodec~\cite{encodec}.}
A convolutional encoder–decoder with RVQ, widely adopted for high-fidelity neural audio compression. We use the official Transformers implementation at 24\,kHz and evaluate standard bandwidths (including 1.5\,kbps) via the model’s built-in encode/decode APIs. In our tables, we report the configured bandwidth along with token/frame statistics measured from emitted code streams.

\paragraph{Mimi~\cite{mimi}.}
A modern low-latency neural codec used in speech–text foundation models. We employ the \textit{kyutai/mimi} Transformers port at 24\,kHz; encoding yields discrete \texttt{audio\_codes} and decoding reconstructs waveforms. Frame/token statistics are computed from code length per second; when codebook details are not exposed, we also report the bandwidth indicated by the model card.

\paragraph{SpeechTokenizer~\cite{speechtokenizer}.}
A unified discrete tokenizer designed for speech LMs (multi-stage VQ with LM-friendly objectives). We load the official package/config and use the provided \texttt{encode}/\texttt{decode} APIs. To reach low-bitrate operating points without retraining, we retain only the first $k$ RVQ layers and optionally apply a time stride; we report the resulting measured token/frame statistics under these settings.

\paragraph{DualCodec~\cite{dualcodec}.}
A dual-branch low-rate codec (semantic + acoustic streams) that balances intelligibility and fidelity. Using the official API at 24\,kHz, we encode both streams and decode with \texttt{decode(semantic, acoustic)}; reported token/frame statistics aggregate both streams.

\paragraph{BigCodec~\cite{bigcodec}.}
A capacity-scaled neural codec targeting very low bitrates with strong reconstruction quality. We call the official repository’s encoder/decoder at 16\,kHz, retaining exposed VQ indices (\texttt{vq\_code}) for rate accounting; any block-size padding mandated by the model is trimmed post-decoding.

\paragraph{TASTE~\cite{taste}.}
A speech tokenizer explicitly designed for joint modeling: text tokens query a frozen speech encoder (keys from the last layer; values from selected shallow layers), producing \emph{text-length} representations that are then discretized before unit decoding and vocoding. We use TASTE as the canonical text-aligned baseline in our low-rate regime.

\paragraph{Ablations used in this paper.}
\textit{S3 Topline:} reconstructs speech directly from ground-truth S3 units with the same unit-to-speech vocoder stack; it serves as an upper-bound reference at word-level frame rates. \textit{Text-only baseline:} predicts S3 tokens from text alone (no speech tokens) to quantify the value of speech tokenization under the same decoder/vocoder. Both are reported in the main results and ablations.

\paragraph{Reproducibility notes.}
For each baseline, we use official or widely adopted checkpoints and libraries (Transformers or the authors’ packages), adhere to their documented sampling rates/hop sizes, and expose only non-retraining knobs (layer keep, time stride) when needed to reach the shared low-bitrate regime. All models are run under the same I/O wrapper described above.

\section{Training Details}
\label{ap:training_details}
We train TASLA on \librispeech{} (train-clean-100/360 and train-other-500) and validate/test on dev-clean/dev-other and test-clean/test-other.
Training uses dynamic batching with a 2000-frame budget and gradient accumulation of 2, distributed across 4 Nvidia A800 GPUs.
The model is initialized from a text-only baseline and employs an FSQ audio quantizer (codebook dimension = 64, codebook size = 8).
We optimize with Adam at a learning rate of \(1.6\times 10^{-4}\) using a warmup scheduler with 5k warmup steps and gradient clipping of 5, for up to 5 epochs.
We evaluate and save every 2000 steps and select the best weights by development-set accuracy.
The random seed is fixed to 1986 for random, NumPy, and PyTorch.

\section{Evaluation Metrics}
\label{ap:metrics}

\subsection{Quality Metrics}
\paragraph{Word Error Rate (WER) \;\(\downarrow\).} 
WER is a standard metric for evaluating speech recognition quality, defined as the minimum edit distance between the hypothesis and reference transcripts, normalized by the reference length. Specifically, it computes the total number of substitutions (\(S\)), insertions (\(I\)), and deletions (\(D\)) required to transform the hypothesis into the reference, divided by the number of words in the reference (\(N\)): 
\[
\mathrm{WER} = \frac{S + D + I}{N}.
\]
A lower WER indicates higher transcription accuracy and better preservation of linguistic content.
We use Whisper-large-v3~\cite{whisper} for transcription and apply standard text normalization before computing WER. This metric reflects intelligibility and alignment with the reference text.

\paragraph{UTMOS \;\(\uparrow\).}
UTMOS~\cite{utmosv2} is a non-intrusive, neural MOS predictor that estimates perceived naturalness and quality directly from the waveform.
Unlike WER, which focuses on content accuracy, UTMOS approximates human subjective ratings of audio quality, providing a complementary measure of perceptual fidelity.

\paragraph{Speaker Similarity \;\(\uparrow\).}
Speaker similarity is quantified as the cosine similarity between speaker embeddings extracted from reference and reconstructed audio.
We use CosyVoice~\cite{cosyvoice} to obtain embeddings.
This metric evaluates whether speaker identity is preserved in the reconstruction.

\subsection{Prosody Metrics}

We report only the prosody metrics used in our tables. Unless noted, audio is resampled to 16\,kHz. Pairwise statistics are computed on a shared, DTW-aligned time axis: we extract MFCCs from reference/hypothesis, obtain a DTW path, and time-warp hypothesis-side prosody features to the reference timeline. Let aligned F0 (Hz) be $f^{\mathrm{Hz}}_{\mathrm{ref}},f^{\mathrm{Hz}}_{\mathrm{hyp}}\in\mathbb{R}^{L}$ with voiced masks $v_{\mathrm{ref}},v_{\mathrm{hyp}}\in\{0,1\}^{L}$. Define the jointly voiced mask $b=v_{\mathrm{ref}}\land v_{\mathrm{hyp}}$ and $n_v=\sum b$. Frame RMS sequences are $\rho_{\mathrm{ref}},\rho_{\mathrm{hyp}}$; their dB versions use $E(\cdot)=20\log_{10}(\max(\cdot,\epsilon))$.

\paragraph{F0 Pearson Correlation (F0-PCC) \;\(\uparrow\).}
Assesses agreement of pitch \emph{contours} (intonation dynamics) while ignoring absolute offsets. Computed as the Pearson correlation between $f^{\mathrm{Hz}}_{\mathrm{ref}}$ and $f^{\mathrm{Hz}}_{\mathrm{hyp}}$ over jointly voiced frames $b_t{=}1$.

\paragraph{Voicing Decision Error (VDE) \;\(\downarrow\).}
Measures consistency of voiced/unvoiced decisions (e.g., missed onsets, spurious voicing). Computed as the framewise disagreement rate
\[
\mathrm{VDE}=\tfrac{1}{L}\sum_{t=1}^{L}\mathbb{1}\!\left[v_{\mathrm{ref}}(t)\neq v_{\mathrm{hyp}}(t)\right].
\]

\paragraph{Gross Pitch Error (GPE) \;\(\downarrow\).}
Counts large pitch mistakes (octave jumps, severe tracking errors) on voiced frames. Let $r(t)=\tfrac{|f^{\mathrm{Hz}}_{\mathrm{hyp}}(t)-f^{\mathrm{Hz}}_{\mathrm{ref}}(t)|}{\max(f^{\mathrm{Hz}}_{\mathrm{ref}}(t),\epsilon)}$ on $b_t{=}1$ and $g(t)=\mathbb{1}[r(t)>0.2]$. Report $\mathrm{GPE}=\tfrac{1}{n_v}\sum_{t:b_t=1} g(t)$.

\paragraph{Energy RMSE (dB) \;\(\downarrow\).}
Captures the mismatch of the loudness envelope (stress/emphasis patterns). Convert frame RMS to dB and compute either RMSE or MSE in the dB domain:
\[
\mathrm{RMSE}_{\mathrm{dB}}=\sqrt{\tfrac{1}{L}\sum_{t}(E(\rho_{\mathrm{ref}}(t))-E(\rho_{\mathrm{hyp}}(t)))^{2}}.
\]
\paragraph{Energy Pearson Correlation (Energy PCC) \;\(\uparrow\).}
Evaluates agreement of dynamic energy contours independent of absolute gain. Compute the Pearson correlation between $E(\rho_{\mathrm{ref}})$ and $E(\rho_{\mathrm{hyp}})$ on the aligned timeline.

\paragraph{Phrase Shape via Legendre Coefficients \;(\(\text{Coeff-}L_2\downarrow\), Coeff-cos\(\uparrow\)).}
Summarizes phrase-level intonation (rise/fall shapes) with low-dimensional, noise-robust descriptors. Convert F0 to semitones $s(t)=12\log_2\!\big(f^{\mathrm{Hz}}(t)/55\big)$, interpolate NaNs on voiced islands, and fit degree-$d$ Legendre polynomials on normalized time $x\in[-1,1]$ to obtain coefficient vectors $\mathbf{c}_{\mathrm{ref}},\mathbf{c}_{\mathrm{hyp}}$. Report
\[
\text{Coeff-}L_2=\lVert\mathbf{c}_{\mathrm{ref}}-\mathbf{c}_{\mathrm{hyp}}\rVert_2,
\]
\[
\text{Coeff-cos}=\frac{\mathbf{c}_{\mathrm{ref}}^{\top}\mathbf{c}_{\mathrm{hyp}}}{\lVert\mathbf{c}_{\mathrm{ref}}\rVert\,\lVert\mathbf{c}_{\mathrm{hyp}}\rVert}.
\]

\paragraph{Implementation notes.}
F0 uses CREPE~\cite{crepe}; energy uses frame RMS; all pairwise metrics share the same DTW-aligned time base to avoid length/misalignment artifacts. Degree 3 for Legendre fits is fixed across systems/datasets for fair comparison.

\section{Visualization of Dynamic Weights}
\begin{figure*}
    \centering
    \includegraphics[width=1\linewidth]{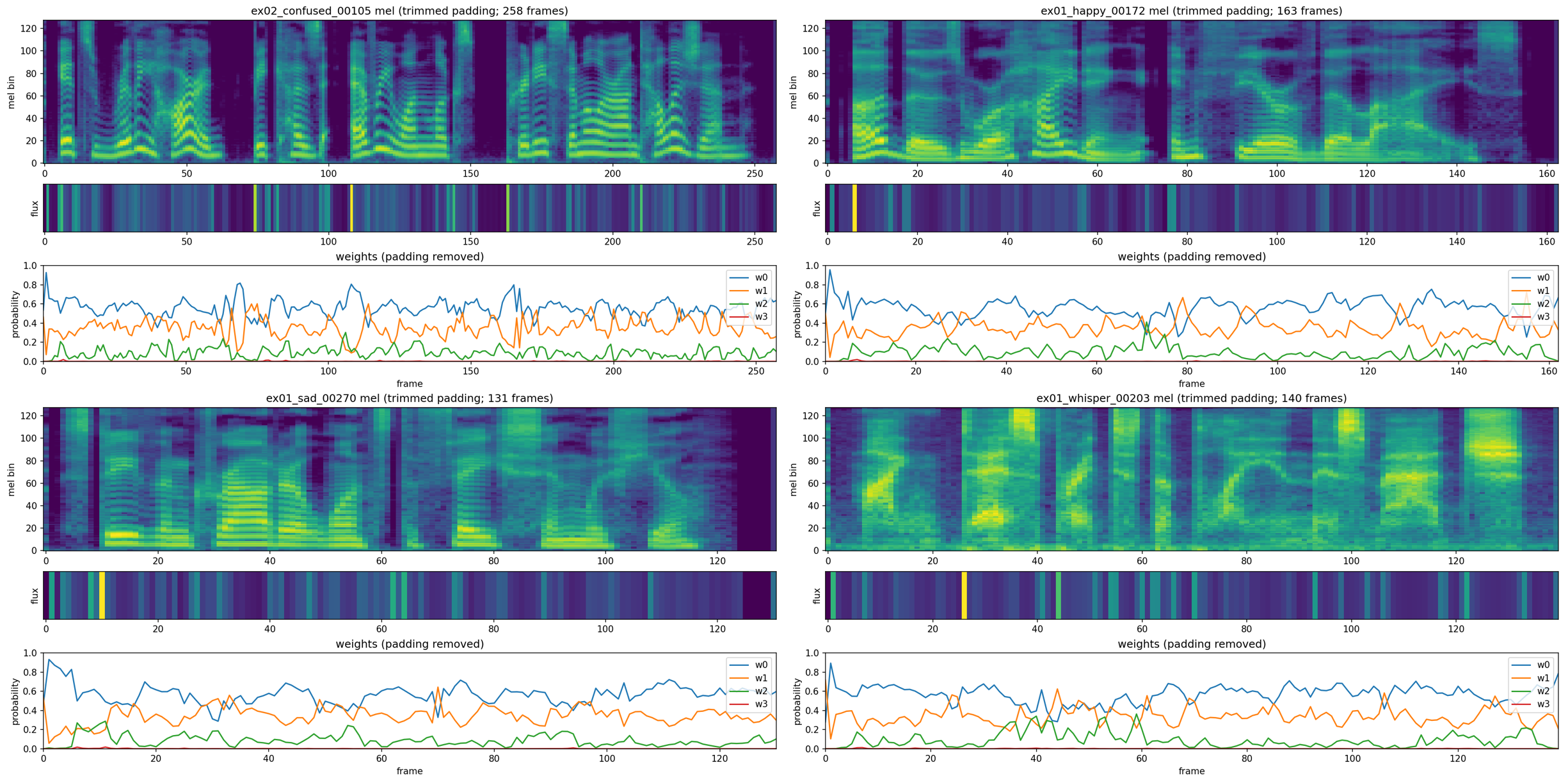}
    \caption{\textbf{Layer-wise Weight Dynamics on \expresso{}.}  
Four examples from \expresso{} showing Mel spectrograms, spectral flux, and dynamic layer weights (\(w_0\)–\(w_3\)).  
\(w_3\) remains consistently small, while \(w_0\), \(w_1\), and \(w_2\) dominate across most frames.
}
    \label{fig:exp}
\end{figure*}

We select four examples from \expresso{} and present their Mel bins, spectral flux, and layer weights (probabilities in the figure).
As shown in \myfig{fig:exp}, \(w_3\) remains consistently small, while \(w_0\), \(w_1\), and \(w_2\) dominate most frames.

\section{Dataset}
\label{ap:dataset}
We select three dataset for evaluation and training.

\paragraph{\librispeech{}.}
\librispeech{}~\cite{librispeech} is a widely used English corpus containing 585 hours of read speech from 2,458 speakers. We use the training and development subsets during training and report evaluation results on the test subset.

\paragraph{\expresso{}.}
\expresso{}~\cite{expresso} is an expressive English speech corpus designed to capture a wide range of emotions and prosodic patterns. We adopt it for out-of-domain evaluation to assess whether our method preserves emotional cues and prosodic richness.

\paragraph{\voxceleb{}.}
\voxceleb{}~\cite{voxceleb} is a large-scale, text-independent speaker identification dataset collected from unconstrained YouTube videos.
Since transcripts are not provided, we do not calculate WER for this dataset.
We include \voxceleb{} as it provides challenging scenarios with varied speakers, spontaneous speaking styles, and background noise, making it suitable for testing robustness and naturalness in speech reconstruction.

All datasets used in this work are publicly released for academic research (\librispeech{} under CC BY 4.0, \voxceleb{} under YouTube TOS, \expresso{} under research license).

\section{S3 Units}
\label{ap:s3-units}
We use \emph{S3 units} as the discrete supervision and the reconstruction target in our pipeline.
Concretely, an external unit recognizer from the CosyVoice~\cite{cosyvoice} stack converts a waveform into a
one-dimensional sequence of discrete unit IDs $s = [s_1,\dots,s_M]$, $s_t \in \{1,\dots,|\mathcal{V}_{\mathrm{S3}}|\}$,
at roughly word-/syllable-level time resolution. These units carry high-level linguistic content along
with coarse prosodic cues and are paired with a unit-to-speech vocoder that maps $s$ back to a waveform.
In TASLA, the unit decoder consumes the text-aligned speech tokens $z_q$ and predicts the S3 sequence
autoregressively (cross-entropy objective), after which the CosyVoice unit-to-speech vocoder reconstructs
audio. We also report an \emph{S3 topline} that bypasses tokenization by feeding ground-truth S3 units
to the same vocoder, providing an informative upper bound under the same unitized reconstruction stack.

\section{Shallow Layer Abaltion}
\label{ap:shallow}
For the shallow-layer ablation study, we select layers 3, 6, 9, and 32 and evaluate the model on \librispeech{}.
As shown in \mytable{tab:shallow}, the shallow-layer ablation achieves performance comparable to \method{} on the PCC (contour) metrics, but it performs significantly worse on Energy RMSE.

\begin{table*}[t]
\centering
\footnotesize
\setlength{\tabcolsep}{4pt}
\renewcommand{\arraystretch}{1.1}
\begin{tabularx}{\linewidth}{l|cYYYYY}
\toprule
\cmidrule(lr){2-7}
Model & Ene. RMSE $\downarrow$ & F0-PCC $\uparrow$ & Phr. Cos. $\uparrow$ & Ene. PCC $\uparrow$& UTMOS $\uparrow$ & WER $\downarrow$ \\
\midrule
S3 Topline            & 6.94 & 0.91 & 0.92 & 0.95 &  3.40 & 0.04 \\
Text-only Baseline    & 10.08 & 0.33 & 0.89 & 0.81 & 3.51 & 0.23 \\
TASTE                 & 8.63 & 0.80 & 0.91 & 0.88 &  3.42 & 0.10 \\
TASLA                 & 6.97 & 0.87 & 0.90 & 0.92 &3.43 & 0.12 \\
TASLA (Shallow Layer) & 7.65 & 0.86 & 0.91 & 0.92 & 3.44 & 0.10 \\
\specialrule{1.2pt}{2pt}{2pt}
\bottomrule
\end{tabularx}
\caption{\textbf{Experimental Results for Shallow Layer Ablation.}
}
\label{tab:shallow}
\end{table*}

\end{document}